\documentclass{nature}
\usepackage{graphicx}
\usepackage{float} 
\makeatletter
\let\saved@includegraphics\includegraphics
\AtBeginDocument{\let\includegraphics\saved@includegraphics}
\renewenvironment*{figure}{\@float{figure}[H]}{\end@float} 
\makeatother

\usepackage{multirow}
\usepackage{subfigure}
\usepackage{mathtools}
\usepackage[colorlinks=true,linkcolor=blue,citecolor=black,urlcolor=blue]{hyperref}
\usepackage{authblk}
\usepackage[labelfont=bf]{caption}
\usepackage[figurename=Figure]{caption}
\usepackage{siunitx}
\usepackage[dvipsnames]{xcolor}
\usepackage{soul}
\usepackage{xfrac}
\usepackage{amsmath}
\usepackage{soul,xcolor}
\usepackage{lineno}
\usepackage{siunitx}

\graphicspath{{./figures/}}

\begin{document}

\setstcolor{red}

\title{Topological magnetism in diluted artificial adatom lattices}
\author[1,2*]{Amal Aldarawsheh}
\author[1,2*]{Samir Lounis}
\affil[1]{Peter Gr\"{u}nberg Institute, Forschungszentrum J\"{u}lich and JARA, D-52425 J\"{u}lich, Germany}
\affil[2]{Faculty of Physics, University of Duisburg-Essen and CENIDE, 47053 Duisburg, Germany} 
\affil[*]{a.aldarawsheh@fz-juelich.de; s.lounis@fz-juelich.de}
\maketitle
\section*{Abstract}
\begin{abstract}

The ability to control matter at the atomic scale has revolutionized our understanding of the physical world, opening doors to unprecedented technological advancements. Quantum technology, which harnesses the unique principles of quantum mechanics, enables us to construct and manipulate atomic structures with extraordinary precision. 

Here, we propose a bottom-up approach to create topological magnetic textures in diluted adatom lattices on the Nb(110) surface. By fine-tuning adatom spacing, previously inaccessible magnetic phases can emerge. Our findings reveal that interactions between magnetic adatoms, mediated by the Nb substrate, foster the formation of unique topological spin textures, such as skyrmions and anti-skyrmions, both ferromagnetic and antiferromagnetic. Since Nb can be  superconducting, our findings present a novel platform with valuable insights into the interplay between topological magnetism and superconductivity, paving the way for broader exploration of topological superconductivity in conjunction with spintronics applications.

\end{abstract}

\section*{Introduction}
The bottom-up construction of artificial nanostructures offers an exceptional framework for investigating synthetic quantum states of matter, meticulously engineered atom by atom~\cite{Khajetoorians2019,Freeney2022}. The groundbreaking creation of the inaugural  quantum corral~\cite{Crommie1993}, enabling the visualization of confined electronic states, propelled scanning tunneling microscopy (STM) and spectroscopy (STS) into indispensable tools for crafting and customizing the electronic and magnetic characteristics of materials at the atomic level~\cite{Manoharan2000, Wiesendanger2009,Franke2011,heinze2011spontaneous,Lounis2014, Delgado2021,Wang2023,Friedrich2024,Taner2024}.

Arranging atoms into chains and clusters facilitates the exploration of a rich array of quantum phenomena, including quantum-confined electrons~\cite{Nilius2002,Stroscio2006,Bouhassoune2014}, Dirac bands~\cite{Yankowitz2012,Park2020}, flat bands~\cite{Slot2017}, and topological defects~\cite{Cheon2015,Drost2017}. The magnetic states hosted by such nanostructures exhibit intriguing complexity influenced by underlying competing interactions~\cite{Hirjibehedin2006,Lounis2007,Lounis2008,Lounis2008b,Mavropoulos2010,Holzberger2013}. Atomic impurities possess the capacity to influence the stability of large spin textures  by either pinning or deflecting them~\cite{fernandes2018universality,Holl2020,Arjana2020,fernandes2020impurity,Reichhardt2022}. Man-engineered nanostructures can induce chiral orbital magnetism~\cite{Brinker2018,Bouaziz2018,fernandes2023} and give rise to novel magnetoresistance effects~\cite{fernandes2020defect}. Superlattices of adatoms can emerge through interactions mediated by surface-state electrons~\cite{Silly2004,Negulyaev2009}, which are influenced by Friedel charge oscillations~\cite{Meier2011,Lounis2012}  leading to long-range magnetic interactions known as Ruderman-Kittel-Kasuya-Yosida (RKKY) interactions~\cite{Ruderman1954,Kasuya1956,Yosida1957}. These interactions, such as the  isotropic Heisenberg interaction and the spin-orbit induced Dzyaloshinskii-Moriya interaction (DMI)~\cite{Dzyaloshinsky1958,moriya1960},  oscillate and decay as a function of distance. They play a dominant role when magnetic atoms are placed directly atop a metal surface and  have been quantified through measurements and electronic simulations in various diluted nanostructures~\cite{Imamura2004,Zhou2010,Khajetoorians2012,Khajetoorians2016,Bouaziz2017,Hermenau2019}.

The capability to fabricate artificial atomic lattices with adjustable inter-atomic distances presents a unique opportunity to explore a vast magnetic phase diagram, a feat unattainable with conventional materials without altering their chemical compositions and structures. One remarkable example that hinges on the delicate balance of various magnetic interactions is the emergence of magnetic skyrmions~\cite{roessler2006spontaneous, fert2013skyrmions,Nagaosa2013,finocchio2016magnetic,Fert2017,everschor2018perspective,zhou2019magnetic,zhang2020skyrmion}.

By adjusting the  separation between magnetic atoms, it becomes possible to toggle the magnetic coupling from FM to AFM, manipulate the chirality governed by the DMI, or even access a regime where the Heisenberg magnetic interaction is eclipsed by the DMI~\cite{Zhou2010,Khajetoorians2012,Khajetoorians2016,Bouaziz2017}. This motivates the design of artificial lattices capable of realizing  topological magnetic textures. Herein, we examine the case of Cr, Mn, or Fe adatoms deposited on a Nb(110) surface, renowned for its superconducting properties and extensively utilized in cutting-edge experiments~\cite{Odobesko2019}. These experiments aim to probe the  potential emergence of topological Majorana boundary states~\cite{Schneider2021,Schneider2022,Kster2022} or trivial ones~\cite{Odobesko2020,Kster2021,Kster2021b,Beck2021,Brinker2022}. Recently, it was demonstrated that two-dimensional diluted lattices comprising Cr adatoms atop a Nb(110) surface host two types of mirror-symmetry-protected topological superconductors~\cite{soldini2023two}.
\begin{figure}
\centering
\includegraphics[width=0.8\linewidth]{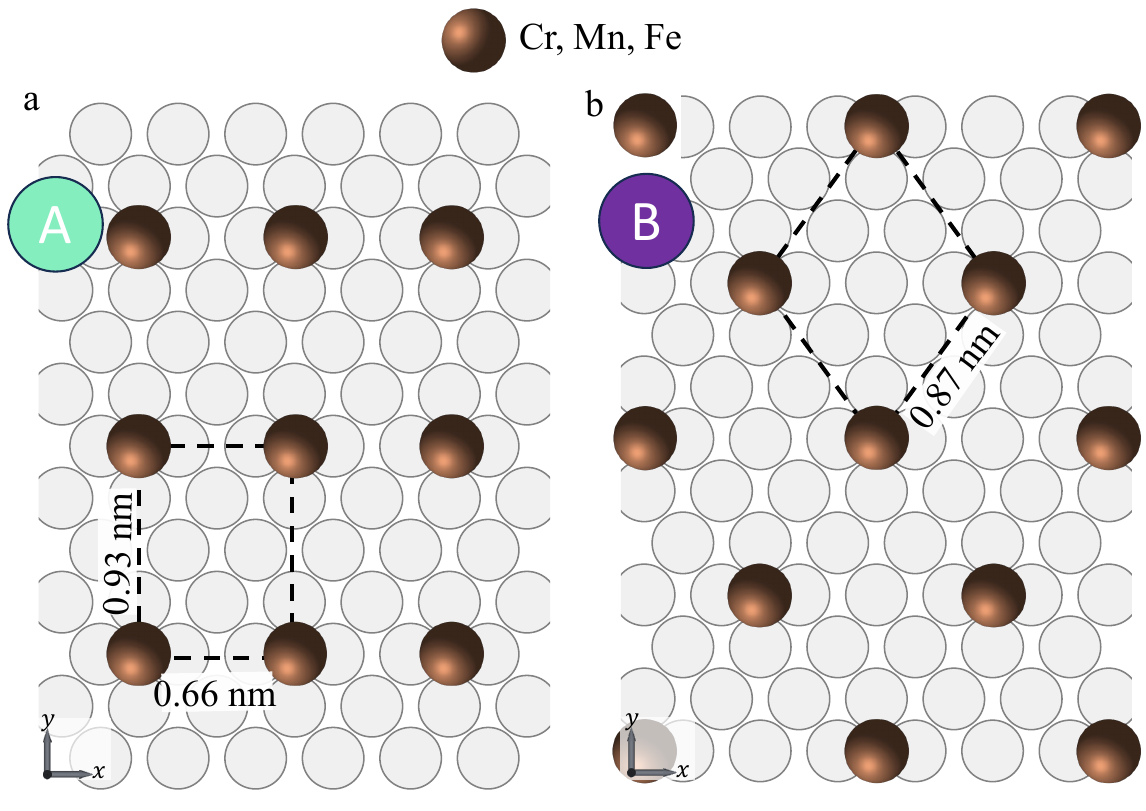}
\caption{\textbf{Diluted lattices.} \textbf{a, b}  Schematic  representation of the  magnetic adatoms positioned in both investigated lattice types, (A) the rectangular lattice with the first (second) nearest neighbours atoms separated by 0.66 nm (0.93 nm), and (B) the rhombic lattice with the first nearest neighbors set 0.87 nm apart. Note that the lattices were built experimentally as reported in Ref.~\cite{soldini2023two}.}
\label{fig:5-1}
\end{figure}
Hinging on first-principles simulations combined with atomic spin-dynamics (ASD) (see computational details section), we unveil the emergence of diverse complex magnetic states such as domain walls, skyrmions, and antiskyrmions through the adatom-adatom magnetic interactions. The explored artificial lattices, inspired from those built experimentally in Ref.~\cite{soldini2023two}, are illustrated in Fig.~\ref{fig:5-1} and denoted as lattices $\mathrm{A}$ and $\mathrm{B}$.  
Noting that skyrmions, being FM or AFM, have been proposed to trigger  the formation of Majorana states once interfaced with a superconductor~\cite{rex2019majorana,diaz2021majorana}, the proposed diluted lattices provide an appealing playground for the exploration of topological superconductivity.

\begin{figure}
\centering
\includegraphics[width=0.8\linewidth]{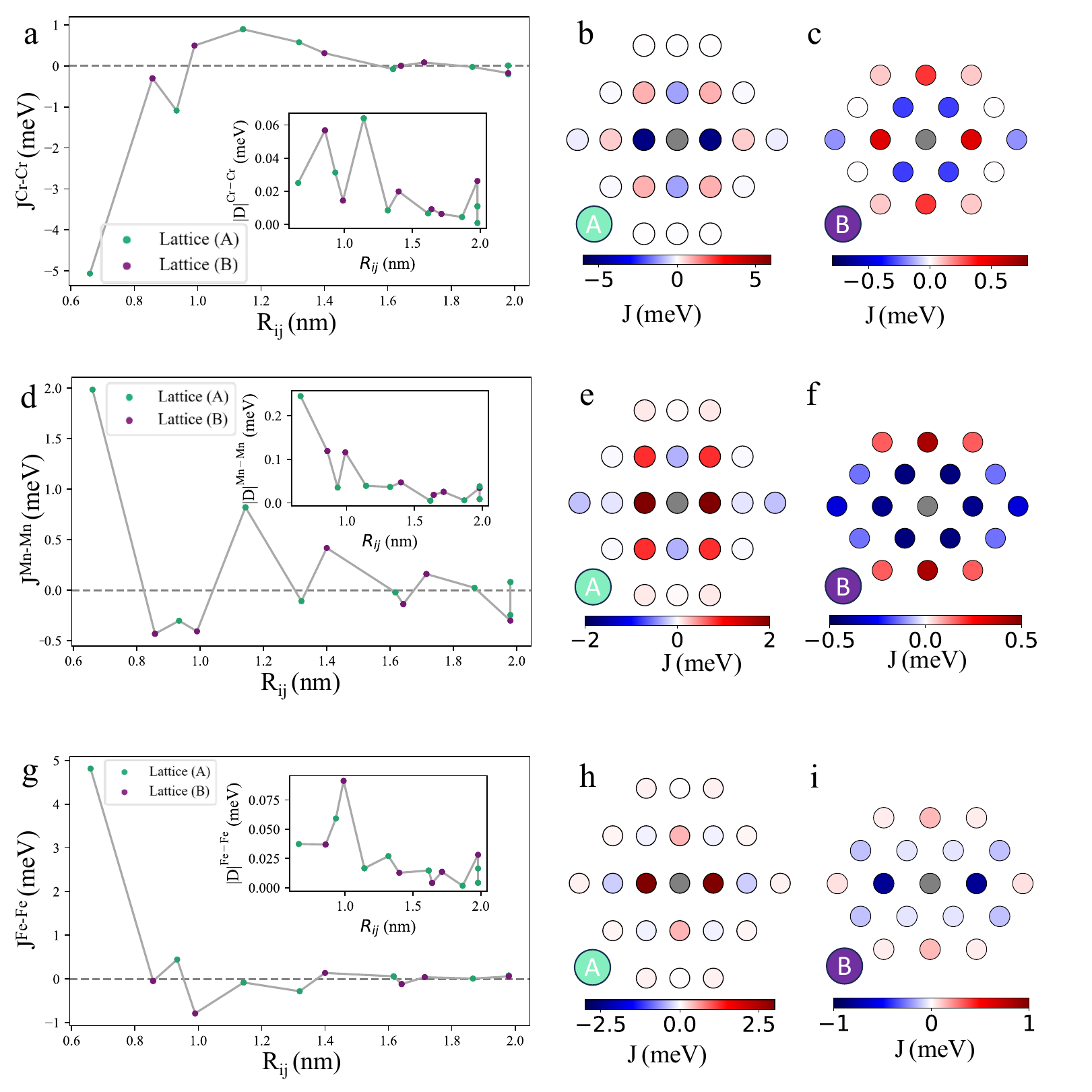}
\caption{\textbf{Magnetic interactions among magnetic adatoms on Nb(110).} \textbf{a, d, g}  Heisenberg exchange interactions among  Cr, Mn, and Fe atoms  as a function of the distance, with the DMI values in inset.  The values obtained for both lattices are plotted in the same curve. The lines serve as a guide for the eye. \textbf{b, c, e, f, h,  i} Illustration of the simulated lattices, where each circle is coloured as a function of the size of J with respect to the central atom (grey colour). The positive (negative) values correspond to FM (AFM) coupling. (Grey lines added for guidance.)}
\label{fig:5-3_new}
\end{figure}

\section{Results}

\subsection{Magnetic interactions  among the magnetic adatoms on Nb(110).}
We start by explore the magnetic interactions among the magnetic adatoms (Cr, Mn, and Fe) shown in 
Fig.~\ref{fig:5-3_new} as function of distance. Our first-principles calculations reveal that the magnetic interactions among the first nearest neighbors (n.n.), and beyond depend strongly on the types of the considered adatom lattices (A or B). For instance, the first n.n.  Heisenberg exchange interactions among Mn adatoms transition from FM coupling ($ \mathrm{J}_{1} $= 2  meV) in the rectangular lattice (lattice (A)) to AFM coupling ($ \mathrm{J}_{1} $= -0.43  meV)  in the rhombic lattice (lattice (B)). For Cr adatoms, the n.n. AFM coupling is reduced by 94\%, decreasing from -5  meV to -0.3  meV. While the Fe adatoms experience a cancellation of the magnetic interaction, initially FM, when placing them in lattice (B) instead of (A). The DMI is found to be finite and can be of the same order of magnitude than the  Heisenberg exchange interactions (e.g. Mn in lattice (B)). Alternatively, in the other cases, it can be one to two orders of magnitude smaller than the Heisenberg exchange interactions.

Notably, not only the Heisenberg exchange interactions and DMI vary across different adatom lattices, but the magnetocrystalline anisotropy energy (MAE) also changes, as shown in Table~\ref{tab:table:8_1}.  The MAE tensor components are defined in the Methods section. 
For Mn adatoms, the MAE exhibits a distinct shift between lattice (A) and lattice (B).  The out-of-plane (OOP) spin orientation is favored in lattice (A) by 0.1 meV with respect to the in-plane magnetization case ($\mathrm{K}_{zz} -\mathrm{K}_{xx}= \mathrm{K}_{zz} -\mathrm{K}_{yy} =  0.1$ meV). Here,  there is no in-plane anisotropy in contrast to all  investigated cases. Conversely,  in lattice (B), there is a transition to  anisotropic spin alignment in the $xz$ plane with ($\mathrm{K}_{yy} - \mathrm{K}_{xx}, \mathrm{K}_{yy}-\mathrm{K}_{zz} )= -$(0.24 meV, 0.18 meV). 

Moving to Cr adatoms lattices, the MAE prefers anisotropic in-plane ($xy$ plane) spin orientation for both lattices, but with different values (($\mathrm{K}_{zz} - \mathrm{K}_{xx}, \mathrm{K}_{zz}-\mathrm{K}_{yy} )= -$(0.31 meV, 0.15 meV), $-$(0.49 meV, 0.21 meV)) for lattice (A), lattice (B), respectively). Similarly,  both  Fe adatoms lattices MAEs prefer anisotropic spin alignment in $yz$ plane with (($\mathrm{K}_{xx} - \mathrm{K}_{yy}, \mathrm{K}_{xx}-\mathrm{K}_{zz} )= -$(0.44 meV, 0.32 meV), $-$(0.58 meV, 1.28 meV)) for lattice (A), lattice (B), respectively). The rich set of MAEs  underscores the tunability of the adatoms  magnetic properties depending on the lattice configurations.

The observed modifications in the magnetic interactions induced by the two types of diluted lattices considered in our study impacts on the ground states and metastable states emerging spin-textures, which are discussed in the next section.
\begin{center}
  \includegraphics[width=1\textwidth]{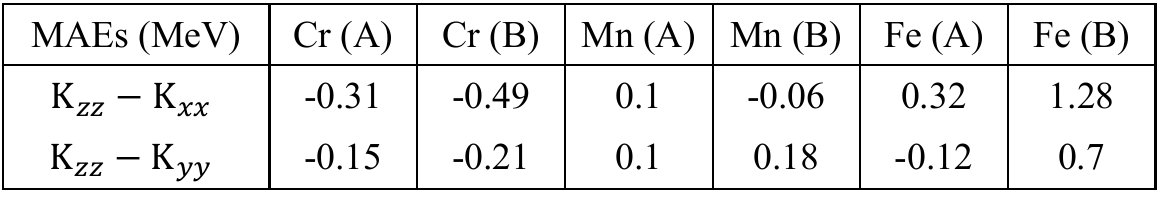}
  \captionof{table}{Tensor elements associated to the  magnetocrystalline anisotropy energies for the different adatoms lattices.}
  \label{tab:table:8_1} 
\end{center}

\subsection{Emerging complex magnetic states. }
After extracting the magnetic interactions among the adatoms for each lattice type, the next step is to investigate the underlying magnetic states.

\begin{figure}
\centering
\includegraphics[width=0.9\linewidth]{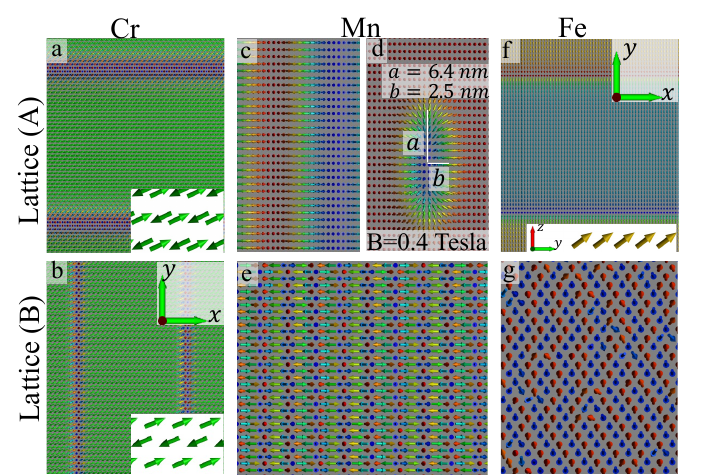}
\caption{\textbf{Magnetic states emerging at Cr, Mn, Fe  lattices.}  \textbf{a, b} Snapshots of the  AFM domain walls  forming on the in-plane ($xy$ plane) AFM ground state  with  the in-plane  anisotropic MAE  ($\mathrm{K}_{zz} - \mathrm{K}_{xx}, \mathrm{K}_{zz}-\mathrm{K}_{yy} )= -$(0.31 meV, 0.15 meV), $-$(0.49 meV, 0.21 meV)) for Cr (A) and (B) lattices, respectively.   For Mn (A) lattice, MAE favors an OOP spin alignment ($\mathrm{K}_{zz} - \mathrm{K}_{xx}, \mathrm{K}_{zz}-\mathrm{K}_{yy} = 0.1$ meV). The magnetic interactions, however, induce FM spin spirals as the ground state \textbf{c}. The application of an OOP magnetic field of 0.4 T enables the stabilization of FM skyrmions \textbf{d}.  \textbf{e} For Mn (B) lattice, MAE prefers anisotropic spin alignment in the $xz$ plane (($\mathrm{K}_{yy} - \mathrm{K}_{xx}, \mathrm{K}_{yy}-\mathrm{K}_{zz} )= -$(0.24 meV, 0.18 meV)), and the ground state  is a complex set of AFM spin spirals. \textbf{f} Snapshot of the FM domains forming at the FM ground state, shown in inset, for the Fe (A) lattice, where the MAE prefers anisotropic spin orientation in the $yz$ plane (($\mathrm{K}_{xx} - \mathrm{K}_{yy}, \mathrm{K}_{xx}-\mathrm{K}_{zz} )= -$(0.44 meV, 0.32 meV)). \textbf{g} For Fe (B) lattice the magnetic interaction among the adatoms, with  an anisotropic MAE in the $yz$ plane (($\mathrm{K}_{xx} - \mathrm{K}_{yy}, \mathrm{K}_{xx}-\mathrm{K}_{zz} )= -$(0.58 meV, 1.28 meV)), gives rise to an irregular AFM order of spins. }
\label{fig:5-15}
\end{figure}

Starting with the case of Cr  adatoms, the ground states are the in-plane row-wise AFM magnetic states for both lattice (A) and (B),  as depicted in the insets of Fig.~\ref{fig:5-15} a, b, where the spins are oriented in-plane due to the underlying in-plane  MAE. 
AFM domain walls emerge across both types of lattices, as shown in Fig.~\ref{fig:5-15} a, b. 

\begin{figure}
\centering
\includegraphics[width=0.9\linewidth]{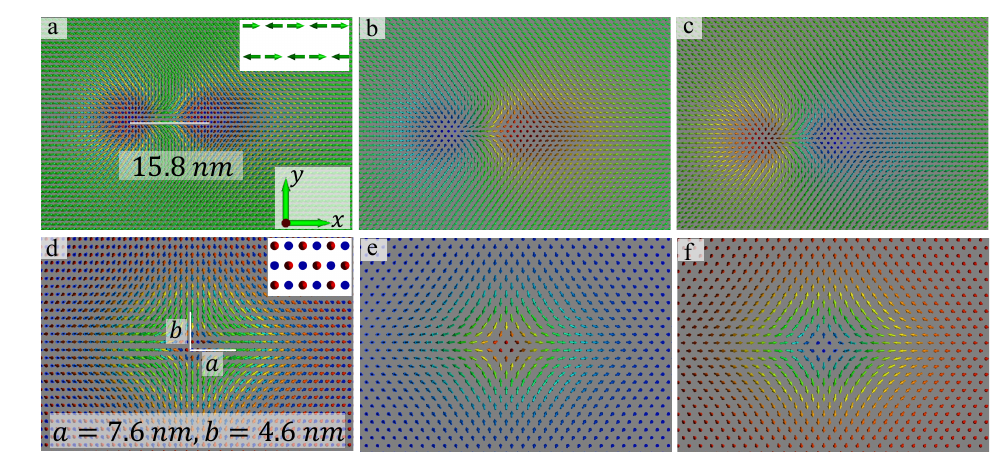}
\caption{\textbf{Magnetic states emerging at Cr (A) lattice by tuning the MAE.}   \textbf{a} Snapshot of the in-plane AFM  Cr antiskyrmion when  $\mathrm{K}_{zz} -\mathrm{K}_{xx}= \mathrm{K}_{zz} -\mathrm{K}_{yy}  = - 0.01$ meV. With this MAE the spins   can align  isotropically in-plane, with the in-plane AFM state shown in inset being the ground state.   \textbf{b, c} The building blocks of the in-plane AFM antiskyrmion, which are two in-plane FM antiskyrmions coupled antiferromagnetically. \textbf{d} Snapshot of the  OOP AFM Cr antiskyrmion after flipping the sign of the MAE ($\mathrm{K}_{zz} -\mathrm{K}_{xx}= \mathrm{K}_{zz} -\mathrm{K}_{yy}  =  0.01$ meV), which leads to an OOP AFM order. \textbf{e, f} The building blocks of the AFM antiskyrmion, which are two FM antiskyrmions coupled antiferromagnetically.}
\label{fig:5-9}
\end{figure}

Since the values of the MAEs are found to be rather small, it is educational to explore the impact of their  magnitude on the magnetic states characterizing the Cr-adatoms-based lattices. When reducing the MAE down to 0.01 meV but  with an isotropic in-plane magnetization ($\mathrm{K}_{zz} -\mathrm{K}_{xx}= \mathrm{K}_{zz} -\mathrm{K}_{yy} = - 0.01$ meV), an in-plane AFM antiskyrmion with a size of 15.8 nm emerges as a metastable state  (Fig.~\ref{fig:5-9} a). Noting that the RW-AFM state can be decomposed into two FM sublattices, which are antiferromagnetically aligned with respect to each other, we can decompose the AFM antiskyrmion into in-plane FM antiskyrmions (Fig.~\ref{fig:5-9} b and c) each residing in one of the FM sublattices. 

Similarly, when flipping the sign of the MAE while keeping the value of 0.01 meV, the OOP AFM configuration is stabilized (see Fig.\ref{fig:5-9} d), which can host an elliptical AFM antiskyrmion. The   major and minor axes of the elliptical AFM antiskyrmion measure 7.6 nm and 4.6 nm, respectively. The constituents of this AFM antiskyrmion are two FM antiskyrmions emerging at FM sublattices, as depicted in Fig.\ref{fig:5-9} e and f. The manifestation of antiskyrmions rather than skyrmions, is due to the DMI vectors that favor a directional dependence of the chirality imposed by the symmetry of the artifical lattices. The chirality is opposite sign along the $x$-direction than that along the $y$-direction.   

\begin{figure}
\centering
\includegraphics[width=0.6\linewidth]{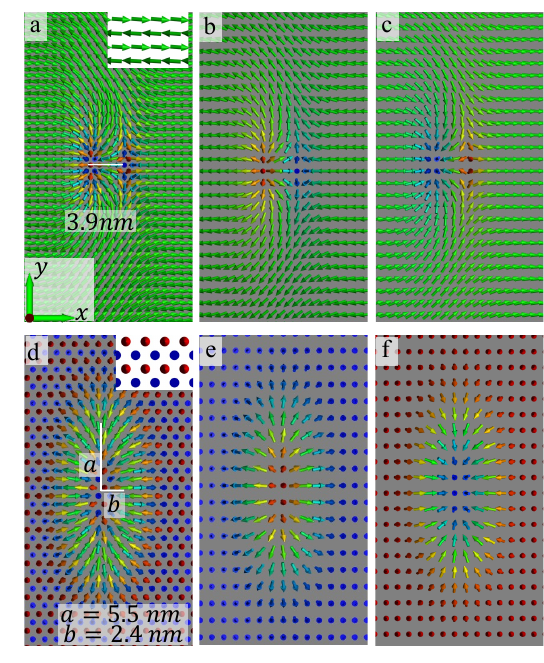}
\caption{\textbf{Magnetic states emerging at Cr (B) lattice by tuning the MAE.} 
\textbf{a} Snapshot depicting the in-plane AFM Cr skyrmion emerging when $\mathrm{K}_{zz} -\mathrm{K}_{xx}= \mathrm{K}_{zz} -\mathrm{K}_{yy}  =  -0.2$ meV. The latter promotes an isotropic in-plane  spin orientation and therefore, the in-plane AFM order is the ground state (inset of \textbf{a}).
\textbf{b, c} The building blocks of the in-plane AFM skyrmion, consisting of two in-plane FM skyrmions that are coupled antiferromagnetically.
\textbf{d} The AFM Cr skyrmion when the MAE favors an OOP spin orientation ($\mathrm{K}_{zz} -\mathrm{K}_{xx}= \mathrm{K}_{zz} -\mathrm{K}_{yy} = 0.08 $ meV), with  the associated OOP AFM ground state shown in inset.  \textbf{e, f} Snapshots of  the constituents   of the AFM skyrmion,  two  FM skyrmions that are antiferromagnetically coupled.}
\label{fig:5-6}
\end{figure}

The same scenario holds for the (B) lattice  of Cr adatoms. Initially, with an isotropic in-plane MAE ($\mathrm{K}_{zz} -\mathrm{K}_{xx}= \mathrm{K}_{zz} -\mathrm{K}_{yy} = - 0.2$ meV), we obtain an in-plane AFM state as depicted in  the inset of Fig.~\ref{fig:5-6} a. In this case, the DMI vectors stabilizes   AFM skyrmions (Fig.~\ref{fig:5-6} a). This  3.9 nm sized in-plane AFM skyrmion is built up of   two in-plane FM skyrmions, residing at two FM sublattices (Fig.~\ref{fig:5-6} b, and c). Whereas upon changing the sign of the MAE ($\mathrm{K}_{zz} -\mathrm{K}_{xx}= \mathrm{K}_{zz} -\mathrm{K}_{yy} = 0.08$ meV)  to favor an OOP spin alignment, the ground state flips from the in-plane orientation to an OOP AFM state depicted in the inset of Fig.~\ref{fig:5-6} d. Here, an OOP AFM skyrmion emerges.  This skyrmion is elliptical in shape with  dimensions of (5.5 nm, 2.4 nm)  as depicted in Fig.~\ref{fig:5-6} d, and the building blocks in this case are two elliptical FM skyrmions residing at two oppositely spin oriented  FM sublattices (Fig.~\ref{fig:5-6} e, and f).

\begin{figure}
\centering
\includegraphics[width=0.9\linewidth]{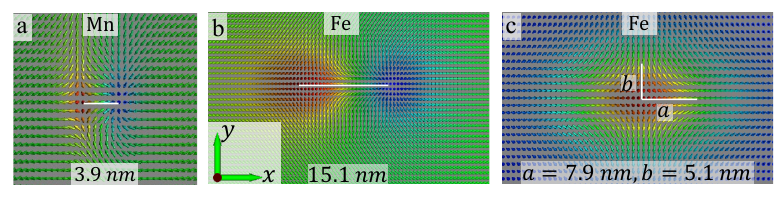}
\caption{\textbf{Magnetic states emerging at Mn and Fe adatoms (A) lattices by tuning MAE.} \textbf{a} Snapshot of the in-plane FM skyrmion forming at Mn (A) lattice when the MAE is preferring in-plane spin alignment  ($\mathrm{K}_{zz} -\mathrm{K}_{xx}= \mathrm{K}_{zz} -\mathrm{K}_{yy} = - 0.5$ meV). \textbf{b} Snapshot of the in-plane FM antiskyrmion  emerging at Fe (A) when the MAE is modified to be isotropic in the $xy$ plane ($\mathrm{K}_{zz} -\mathrm{K}_{xx}= \mathrm{K}_{zz} -\mathrm{K}_{yy}  = -0.025$ meV).  Flipping the sign of the MAE ($\mathrm{K}_{zz} -\mathrm{K}_{xx}= \mathrm{K}_{zz} -\mathrm{K}_{yy}  = 0.015$ meV) favors the OOP spin orientation, where the ground state is the FM state,  which hosts  FM antiskyrmions \textbf{c}.  } 
\label{fig:5-8}
\end{figure}

Regarding the Mn-based (A) lattice case, the ground state is FM spin spirals in the absence of magnetic filed, as shown in Fig.~\ref{fig:5-15} c. Upon applying an OOP magnetic field of 0.4 Tesla, the spin spirals deform into elliptical FM skyrmions surrounded by an OOP FM state (Fig.~\ref{fig:5-15} d). 
 In this case the ellipse has major and minor axes of 6.4 nm, and 2.5 nm.     By flipping the sign of the MAE, and choosing it to prefer in-plane spin alignment ($\mathrm{K}_{zz} -\mathrm{K}_{xx}= \mathrm{K}_{zz} -\mathrm{K}_{yy} = - 0.5$ meV), a 3.9 nm sized in-plane FM skyrmion emerges at the in-plane FM background (Fig.~\ref{fig:5-8} a). For Mn-based (B) lattices, the magnetic interactions among the adatoms do not support the stabilization of topological solitons, and only complex sets of AFM  spin spirals  emerge as the ground state,   see Fig.~\ref{fig:5-15} e. 

Finally,  for the Fe adatoms lattices, the FM state in the $yz$ plane is the ground state  for the (A) lattice, which can host FM magnetic domain walls (see Fig.~\ref{fig:5-15} f and its inset).  Upon reducing the MAE value down to 0.025 meV, and choosing it to be preferring an isotropic in-plane ($xy$ plane) orientation of spins,  i.e. $\mathrm{K}_{zz} -\mathrm{K}_{xx}= \mathrm{K}_{zz} -\mathrm{K}_{yy} = - 0.5$ meV,  a 15.1 nm sized in-plane FM antiskyrmion emerges (Fig.~\ref{fig:5-8} b). Whereas, when flipping the sign of the MAE, 
 an elliptical FM antiskyrmion emerges in the OOP FM background as shown in Fig.~\ref{fig:5-8} c, with major and minor axes of 7.9 nm, 5.1 nm, respectively. For the Fe (B) lattices, the weak magnetic interactions among the  Fe adatoms ($\mathrm{J}_{1}= -0.03$ meV)  do not support the stabilization of topological solitons, and only AFM irregular (kind of arbitrary) spin alignments    emerge, see Fig.~\ref{fig:5-15} g.

 To summarize, we have learned from this study that Mn-based dilute lattice (A) is the most promising case to explore the emergence of topological magnetic states. For the other cases, the complexity of the MAE tensor, which shows biaxial anistropy, breaks magnetic rotation in the plane encompassing the spin-textures. This  works against the formation of solitonic spin-textures.  Restoring an isotropic in-plane rotation enables the formation of various magnetic objects such as AFM of FM skyrmions or antiskyrmions.

\section{Discussion }

In this study, we unveiled the emergence of a plethora of topological solitons on the diluted lattices of Cr, Mn, and Fe adatoms deposited on Nb(110) surface. We find potential stabilization of  FM and AFM skyrmions, as well as antiskyrmions, which depend on the lattice type of the adatom structures. Motivated by recent STM experiments,  demonstrating the possibility of creating diluted lattices, we assumed either a rectangular lattice (lattice (A)), or a rhombic lattice (lattice (B)).

The Heisenberg  exchange interactions, DMI and MAE can be strongly modified depending on the lattice considered. For instance, the coupling between the n.n. adatoms can change from being FM to AFM such as what we observed for the case of  Mn adatoms. Moreover the DMI chirality changes across the different lattices types stabilizing for example antiskyrmions in the (A) lattices of Cr and Fe adatoms, while in the Mn (A) and Cr (B) lattices,  skyrmions are formed.

In conclusion, our study provides a comprehensive understanding of the magnetic interactions and topological spin textures in diluted adatom lattices. This research opens up new possibilities for exploring synthetic quantum states of matter and their potential applications in  technologies. Our investigations promote the superconducting substrate Nb for the exploration of topological superconducting states via the emergence of Majorana states expected to accompany skyrmionic states. 
Future research could focus on the diluted lattices on several interfaces such as Ir(111), which might lead to more isotropic MAE.

\begin{methods}

\label{ch:paper5_computaional}

We conducted a systematic investigation to explore the magnetic structures that can be hosted by  the magnetic diluted lattices in our six  layered systems. Our approach involves a two-fold procedure, combining \textit{ab initio} calculations with ASD. Similarly to the experimental construction of the adatoms-based diluted structures reported in Ref.~\cite{soldini2023two}, we consider two possible lattices denoted as ($\mathrm{A}$) and ($\mathrm{B}$), visually depicted in Fig.~\ref{fig:5-1}. We assume a  slab configuration consisting of 5 Nb layers and 1 diluted adatoms-based layer. 
In each layer we have 8 atoms per unit cell for lattice $\mathrm{A}$, and 9 atoms per unit cell for system $\mathrm{B}$. We place the diluted magnetic layer (the magnetic adatoms are either Cr, Mn or Fe) such that the adatoms reside on the hollow stacking site as depicted in Fig.~\ref{fig:5-1}.

In the first step, we conducted detailed investigation of the magnetic properties and interactions for the different systems, using the all-electron full-potential relativistic Korringa-Kohn-Rostoker  (KKR) Green function method, implemented in the JuKKR computational package~\cite{papanikolaou2002conceptual,bauer2014development},in the local spin density approximation. To perform the calculations, the momentum expansion of the Green function was truncated at $\ell_{\text{max}} = 3$. Self-consistent calculations were conducted using a k-mesh of $30\times21\times1$ points for lattice $\mathrm{A}$, and of  $30\times30\times1$ points for lattice $\mathrm{B}$ . The energy contour consisted of 32 complex energy points in the upper complex plane, and it incorporated 10 Matsubara  poles.
To extract the Heisenberg exchange interactions and Dzyaloshinskii-Moriya (DM) vectors ~\cite{Dzyaloshinsky1958,moriya1960,yang2023first}, we employed the infinitesimal rotation method ~\cite{ Liechtenstein1987,Ebert2009}. For this extraction, we used a finer k-mesh of $300\times210\times1$ points for lattice   $\mathrm{A}$, and of  $300\times300\times1$ points for lattice $\mathrm{B}$.

After extracting the magnetic interactions characterizing the adatoms, we solve,  the Landau-Lifshitz-Gilbert (LLG) equation to  minimize the underlying extended Heisenberg Hamiltonian:

\begin{align}
   H &= H_\text{Exchange} + H_\text{DMI} + H_\text{Anisotropy} + H_\text{Zeeman} \notag \\
   &= -\sum\limits_{<i,j>} J_{ij} \textbf{S}_{i}\cdot \textbf{S}_{j}
   - \sum\limits_{<i,j>} \mathbf{D}_{ij} \cdot (\mathbf{S}_{i} \times \mathbf{S}_{j}) 
   -  \sum\limits_{i}  \mathbf{S}_{i}^{T} \mathcal{K}_{i}  \mathbf{S}_{i} 
   - \sum\limits_{i} \mu_i \mathbf{B} \cdot \mathbf{S}_i
   \label{eq.Heisenberg}
\end{align}

where  we assign indices $i$ and $j$ to denote specific sites, each associated with a magnetic moment. $\text{X}$ represents the magnetic atoms being Cr, Mn or Fe adatoms. The magnetic moment is represented by the unit vector $\textbf{S}$.  $J$ is the Heisenberg exchange coupling strength, being negative for an AFM interaction

Similarly, we use the notation $\textbf{ D}$ for the Dzyaloshinskii-Moriya interaction vector,  $\mu_i\textbf{B}$ to represent the Zeeman coupling to the atomic spin moment $\mu$ at site $i$, and  $\mathcal{K}$ is  the magnetocrystalline anisotropy energy tensor, whose components are  calculated from the energy differences between different magnetic orientations.  Taking the energy difference between the cases where the magnetization points along the $x$ and $z$ axis gives:
$\varepsilon^x-\varepsilon^z = \mathrm{K}_{zz} - \mathrm{K}_{xx}$, while considering the magnetization along the $x$ and $y$ axis leads to:  $\varepsilon^x-\varepsilon^y = \mathrm{K}_{yy} - \mathrm{K}_{xx}.$

To explore the magnetic properties and emerging complex states we utilize the Landau-Lifshitz-equation  (LLG) as implemented in the Spirit code~\cite{muller2019spirit}. We assumed periodic boundary conditions to model the extended two-dimensional system with cells containing $100^2$, $200^2$, and $300^2$ sites.

\end{methods}

\section*{References}
\bibliographystyle{naturemag}
\bibliography{references}

\end{document}